\documentclass{aa}

\newcommand\dosingle[1]{#1}  \newcommand\dodouble[1]{ } 

\newcommand\nice[1]{#1}    \newcommand\subm[1]{}   
\subm{ \renewcommand\dosingle[1]{ }   \renewcommand\dodouble[1]{#1} }

\dodouble{ \documentclass[referee]{aa} }

\usepackage{graphicx}

\usepackage{amsfonts}
\usepackage{url}

\newcommand\prerefereechanges[1]{#1}  \newcommand\prerefereestart{  }  \newcommand\prerefereestop{ }

     \usepackage{color}  \definecolor{myred}{rgb}{0.7,0.0,0.2} 

 \usepackage{hyperref}

\nice{ 
\providecommand{\eprint}[1]{\href{http://arxiv.org/abs/#1}{{\tt [arXiv:#1]}}}

\providecommand{\url}[1]{\href{#1}{#1}}

}

\providecommand{\adsurl}[1]{} 
\newcommand\SSS{Sect.~}

\nice{ \newcommand\mycaptionfont{\protect\footnotesize} }

\newcommand\gtapprox{\,\lower.6ex\hbox{$\buildrel >\over \sim$} \, }
\newcommand\ltapprox{\,\lower.6ex\hbox{$\buildrel <\over \sim$} \, }
\newcommand\mapstoapprox{\,\lower.6ex\hbox{$\buildrel \mapsto \over \sim$} \, }
\newcommand\propapprox{\,\lower.6ex\hbox{$\buildrel \propto\over \sim$} \, }

\newcommand\arcs{\ifmmode {'' }\else $'' $\fi}     
\newcommand\arcm{\ifmmode {' }\else $' $\fi}       
\newcommand\ddeg{\ifmmode^\circ\else$^\circ$\fi}    

\newcommand\frtoday{Le\space\number\day\space\ifcase\month\or
  janvier\or f\'evrier\or mars\or avril\or mai\or juin\or
  juillet\or ao\^ut\or septembre\or octobre\or novembre\or 
d\'ecembre\fi\space \number\year}
\newcommand\todayISO{\number\year-\ifnum\month<10 0\fi\number\month-\ifnum\day<10 0\fi\number\day}

\newcommand\gh{g_{\mathrm{h}}} 
\newcommand\Npix{N_{\mathrm{pix}}} 
\newcommand\Ntod{N_{\mathrm{TOD}}} 
\newcommand\xim{x_{\mathrm{im}}} 
\newcommand\muone{\left.\left<\delta T\right>\right|_{\phi,\theta}} 
\newcommand\mutwo{\left.\left<(\delta T)^2\right>\right|_{\phi,\theta}} 
\newcommand\sigperpix{\sigma_{g_{\delta t}^{-1}(\delta T)}\big|_{\phi,\theta}}
\newcommand\mapdipole{m_{\mathrm{0}}}
\newcommand\mapannualdipole{M_{\mathrm{1}}}

\newcommand\notea{^\mathrm{a}}
\newcommand\noteb{^\mathrm{b}}
\newcommand\notec{^\mathrm{c}}
\newcommand\noted{^\mathrm{d}}
\newcommand\notee{^\mathrm{e}}

\title{On the suspected timing-offset--induced calibration error in the Wilkinson microwave anisotropy probe time-ordered data}

\author{Boudewijn F. Roukema
}

\institute{Toru\'n Centre for Astronomy, Nicolaus Copernicus University,
ul. Gagarina 11, 87-100 Toru\'n, Poland 
}

\date{\frtoday}

\titlerunning{Suspected WMAP TOD calibration error}
\authorrunning{Roukema}

\begin{document}


\newcommand\Nchainsmain{16}
\newcommand\Npergroup{four}

\abstract
{In the time-ordered data (TOD) files of the Wilkinson Microwave
  Anisotropy Probe (WMAP) observations of the cosmic microwave
  background (CMB), there is 
  an undocumented timing offset of $-25.6$~ms between
  the spacecraft attitude and radio flux density timestamps
  in the Meta and Science Data Tables, respectively.
  If the offset induced an error during calibration of the raw
  TOD, then estimates of the WMAP CMB quadrupole might be
  substantially in error.}
{A timing error during calibration would not only induce an artificial
  quadrupole-like signal in the mean sky map, it would also add variance
  per pixel. This variance would be present in the calibrated TOD.
  Low-resolution map-making as a function of timing offset should show
  a minimum variance for the correct timing offset.}
{Three years of the calibrated, filtered WMAP 3-year TOD are compiled
  into sky maps at HEALPix resolution $N_{\mathrm{side}}=8$,
  individually for each of the K, Ka, Q, V and W band differencing
  assemblies (DA's), as a function of timing offset. The median 
  per map of the temperature fluctuation variance per pixel is calculated
  and minimised against timing {\protect\prerefereechanges{offset}}.}
{Minima are clearly present. The timing offsets that minimise the
  median variance are \protect\prerefereechanges{{$-38\pm8$~ms (K,
      Ka), $-27\pm3$~ms (Q), $-43\pm8$~ms (V), and $-47\pm194$~ms (W),
      i.e.  an average of $-30\pm 3$~ms, where the WMAP
      collaboration's preferred offset is $0\pm1.7$~ms. A
      non-parametric bootstrap analysis rejects the latter at a
      significance of $99.999\%$.  The hypothesis of a $-25.6$~ms
      offset, suggested by Liu, Xiong \& Li from the TOD file timing
      offset, is consistent with these minima.}}}
{It is difficult to avoid the conclusion that the WMAP calibrated TOD
  and inferred maps are wrongly calibrated.  CMB quadrupole estimates
  $(3/\pi)C_2$ based on the incorrectly calibrated TOD are
  overestimated by roughly \protect\prerefereechanges{$64\pm6\%$ (KQ85 mask) to $94\pm10\%$} (KQ75
  mask).  Ideally, the WMAP map-making pipelines should be redone
  starting from the {{\em uncalibrated}} TOD and using the $-25.6$~ms
  timing offset correction.}


\maketitle

\dodouble{ \clearpage } 



\newcommand\thypothesis{
\begin{table} 
\caption{\mycaptionfont Hypotheses of timing offset $\delta t$
expressed in 
units of an observing interval $\Delta t$, and in ms.
\label{t-hypothesis}}
$$\begin{array}{c c c c cc } \hline
\rule[-1.5ex]{0ex}{4.5ex}
&   
& \delta t& \sigma_{\delta t }{\notea} &
\rule{1.0ex}{0ex} 
(\delta t -0.5)\Delta t 
\rule{1.0ex}{0ex} 
 & \sigma_{\delta t \Delta t}{\notea} \\
\mathrm{units:} & & \Delta t & \Delta t & \mathrm{ms} & \mathrm{ms} \\
\mathrm{hypothesis} &
\mathrm{band}  \\
\hline
\mathrm{WMAP}{\noteb} 
\rule[-1.5ex]{0ex}{4.5ex} &
\mbox{all} & 0.5 & -{\notec} & ~~~0.0 & 1.7 \\
\hline
 \rule[-1.5ex]{0ex}{4.5ex} 
\mbox{LXL2010 (i)}{\noted} &
\mbox{K, Ka} &    0 &    0.01 & -64.0&   1.7 \\
& \mathrm{Q}   &    0 &  0.02 &  -51.2 &  1.7 \\
& \mathrm{V}   &    0 & 0.02 &   -38.4&   1.7 \\
& \mathrm{W}   &    0 & 0.03 &  -25.6 &  1.7 \\
\hline 
\rule[-1.5ex]{0ex}{4.5ex} 
\mbox{LXL2010 (ii)}{\notee} &
\mbox{K, Ka} &    0.30 & 0.01&   -25.6 &  1.7 \\
& \mathrm{Q}   &    0.25 & 0.02&   -25.6 &  1.7 \\
& \mathrm{V}   &    0.17 & 0.02 &  -25.6 &  1.7 \\
& \mathrm{W}   &    0.00 & 0.03 &  -25.6 &  1.7 \\
\hline
\end{array}$$
\\
${}{\notea}$\protect\nocite{Bennett03MAP}{Bennett} {et~al.} (\protect\hyperlink{hypertarget:Bennett03MAP}{2003a})'s ({\protect\SSS}{6.1.2}) estimate of the
``relative accuracy [that] can be achieved between the star tracker(s), gyro and
the instrument'' is adopted here for each of the hypotheses.
\\ 
${}{\noteb}$WMAP collaboration's preferred timing offset.
\\
${}{\notec}$Same per waveband as for \protect\nocite{LL10toffset}{Liu} {et~al.} (\protect\hyperlink{hypertarget:LL10toffset}{2010a}) hypotheses.
\\
${}{\noted}$\protect\nocite{LL10toffset}{Liu} {et~al.} (\protect\hyperlink{hypertarget:LL10toffset}{2010a})'s hypothesis (i), i.e. described in terms
of the beginning of an observing time interval.
\\
${}{\notee}$\protect\nocite{LL10toffset}{Liu} {et~al.} (\protect\hyperlink{hypertarget:LL10toffset}{2010a})'s hypothesis (ii), i.e. the TOD file
timing offset.   
\end{table}
}  

\newcommand\tKQVW{
\begin{table}
\caption{\mycaptionfont Quadratic fit estimates
of the timing offset
[median 
$\mu_{1/2}(\tau_i )$
and standard error in the median
$\sigma_{\mu_{1/2}(\tau_i)}$, Eq.~(\protect\ref{e-defn-tau})] 
that minimises the standard deviation per pixel
\protect\prerefereechanges{over $\delta t$ intervals symmetric around
$\delta t = 0.5$ in the range $-4 \le \delta t \le 5$,}
for 
6 DA/year combinations $i$ in the K and Ka bands grouped together, 
for 
6 DA/year combinations $i$ in each of the Q and V bands separately, and  
for 12 DA/year combinations $i$ in the W band, shown in 
units of an observing interval $\Delta t$, and in ms, compared
to the WMAP collaboration preferred value of $\delta t=0.5$.
\label{t-KQVW}}
$$\begin{array}{c rr r rr } \hline
\rule[-1.5ex]{0ex}{4.5ex}
&   
\mu_{1/2}( \tau_i )& 
\sigma_{\mu_{1/2}( \tau_i )}& 
\Delta t &
\mu_{1/2}( \tau_i-0.5 ) \Delta t& 
\sigma_{\mu_{1/2}( \tau_i )} \Delta t  \\
\mathrm{units:} & \Delta t  & \Delta t & \mathrm{ms} & \mathrm{ms} & \mathrm{ms} \\
\mathrm{band}  \\
\hline
\rule[-1.5ex]{0ex}{4.5ex} 
\mbox{K, Ka} & 0.20 &   0.06 &  128.0  & -38.4   &    8.3 \\ 
\mathrm{Q} &   0.24 &   0.03 &  104.2  & -27.1   &    2.8 \\ 
\mathrm{V} &     -0.06 &   0.11 &   76.8  & -43.0   &    8.1 \\ 
\mathrm{W}^{\notea} & -0.41 &   3.74 &   52.1  & -47.4   &  194.8 \\ 
\hline
\rule[-1.5ex]{0ex}{4.5ex} 
\mathrm{all}^{\noteb} & 0.22 & 0.02 & - & -29.7 & 2.5 \\
\hline
\end{array}$$
\\
${}{\notea}$In W, the best-fit parabola for
one of the year/DA combinations had a maximum and was ignored.
\\
${}{\noteb}$Mean and standard error, weighted by the standard error,
calculated either in observing interval units (ignoring the 
dependence of $\Delta t$ on waveband), or in ms compared 
to $\delta t=0.5$.
\\
\end{table}
}  

\newcommand\tdomainshift{
\begin{table}
\caption{\mycaptionfont 
Quadratic fit estimates
[median $\mu_{1/2}( \tau_i-0.5 ) \Delta t$ and 
and standard error in the median, in ms]
of minimum standard deviation,
as for Table~\protect\ref{t-KQVW}, using 
\protect\prerefereechanges{$\delta t$ intervals 
symmetric around $\delta t = 0.0$
in the range $-5 \le \delta t \le 5$, i.e. possibly 
showing bias against $\delta t = 0.5$. }
\label{t-domainshift}}
$$\begin{array}{c rr r rr } \hline
\rule[-1.5ex]{0ex}{4.5ex}
&   
\mu_{1/2}( \tau_i )& 
\sigma_{\mu_{1/2}( \tau_i )}& 
\Delta t &
\mu_{1/2}( \tau_i-0.5 ) \Delta t& 
\sigma_{\mu_{1/2}( \tau_i )} \Delta t  \\
\mathrm{units:} & \Delta t  & \Delta t & \mathrm{ms} & \mathrm{ms} & \mathrm{ms} \\
\mathrm{band} \\
\hline
\rule[-1.5ex]{0ex}{4.5ex} 
\mbox{K, Ka} &   0.20 &   0.07 &  128.0  & -37.9   &    9.2 \\ 
\mathrm{Q}   &   0.21 &   0.04 &  104.2  & -30.3   &    3.7 \\ 
\mathrm{V}   &  0.15 &   0.13 &   76.8  & -26.7   &    9.8 \\ 
\mathrm{W}^{\notea}   &   -0.17 &  11.56 &   52.1  & -35.0   &  602.3 \\ 
\hline
\rule[-1.5ex]{0ex}{4.5ex} 
\mathrm{all} & 0.21 & 0.03 & - & -30.8 & 3.2  \\
\hline
\end{array}$$
${}{\notea}$In W, the best-fit parabola for
one of the year/DA combinations had a maximum and was ignored.
\\
\end{table}
}  

\newcommand\teighty{
\begin{table}
\caption{\mycaptionfont 
\protect\prerefereechanges{Quadratic fit estimates for 
usage of the calibrated TOD with an extra term, i.e. using the 80-th map iteration,
as for Table~\protect\ref{t-KQVW}, for the bands with shorter
calculation times.}
\label{t-eighty}}
$$\begin{array}{c rr r rr } \hline
\rule[-1.5ex]{0ex}{4.5ex}
&   
\mu_{1/2}( \tau_i )& 
\sigma_{\mu_{1/2}( \tau_i )}& 
\Delta t &
\mu_{1/2}( \tau_i-0.5 ) \Delta t& 
\sigma_{\mu_{1/2}( \tau_i )} \Delta t  \\
\mathrm{units:} & \Delta t  & \Delta t & \mathrm{ms} & \mathrm{ms} & \mathrm{ms} \\
\mathrm{band} \\
\hline
\rule[-1.5ex]{0ex}{4.5ex} 
\mbox{K, Ka} &   0.32 &   0.06 &  128.0  & -23.3   &    7.3 \\ 
\mathrm{Q}   &   0.40 &   0.13 &  104.2  & -10.0   &   13.7 \\ 
\mathrm{V}   &   0.11 &   0.65 &   76.8  & -30.3   &   49.7 \\ 
\hline
\rule[-1.5ex]{0ex}{4.5ex} 
\mathrm{all} & 
0.33 & 0.05 & - & -20.5 & 6.4  \\
\hline
\end{array}$$
\end{table}
}  

\newcommand\trms{
\begin{table}
\caption{\mycaptionfont 
\protect\prerefereechanges{Quadratic fit estimates
of minimum rms, 
as for Table~\protect\ref{t-KQVW}.}
\label{t-rms}}
$$\begin{array}{c rr r rr } \hline
\rule[-1.5ex]{0ex}{4.5ex}
&   
\mu_{1/2}( \tau_i )& 
\sigma_{\mu_{1/2}( \tau_i )}& 
\Delta t &
\mu_{1/2}( \tau_i-0.5 ) \Delta t& 
\sigma_{\mu_{1/2}( \tau_i )} \Delta t  \\
\mathrm{units:} & \Delta t  & \Delta t & \mathrm{ms} & \mathrm{ms} & \mathrm{ms} \\
\mathrm{band} \\
\hline
\rule[-1.5ex]{0ex}{4.5ex} 
\mbox{K, Ka} &   0.49 &   0.13 &  128.0  &  -1.6   &   16.9 \\ 
\mathrm{Q}   &  0.26 &   0.04 &  104.2  & -25.1   &    4.4 \\ 
\mathrm{V}   &  0.05 &   0.12 &   76.8  & -34.2   &    9.2 \\
\mathrm{W}   &   -0.27 &   0.52 &   52.1  & -40.1   &   27.3 \\
\hline
\rule[-1.5ex]{0ex}{4.5ex} 
\mathrm{all} & 0.25 & 0.04 & - & -25.8 & 3.9  \\
\hline
\end{array}$$
\end{table}
}  

\newcommand\ftoy{
\begin{figure}
\centering 
\includegraphics[width=8cm]{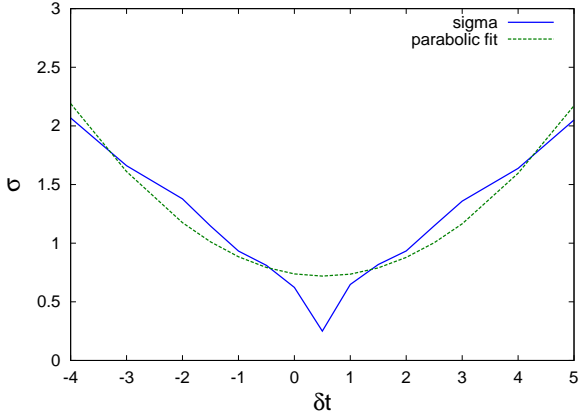}
\caption[]{ \mycaptionfont 
\protect\prerefereechanges{Minimisation 
of a median-per-map standard deviation per pixel
$\sigma(\delta t)$ [Eq.~(\protect\ref{e-sigperpix})]
in a realisation of a toy simulation of would-be wrongly calibrated data with an 
input timing offset of $\delta t = 0.5$, fit by a parabola over a range of timing
offsets centred at $\delta t = 0.5$.
\label{f-toy}}}
\end{figure} 
} 

\newcommand\fKQVW{
\begin{figure}
\centering 
\includegraphics[width=7cm]{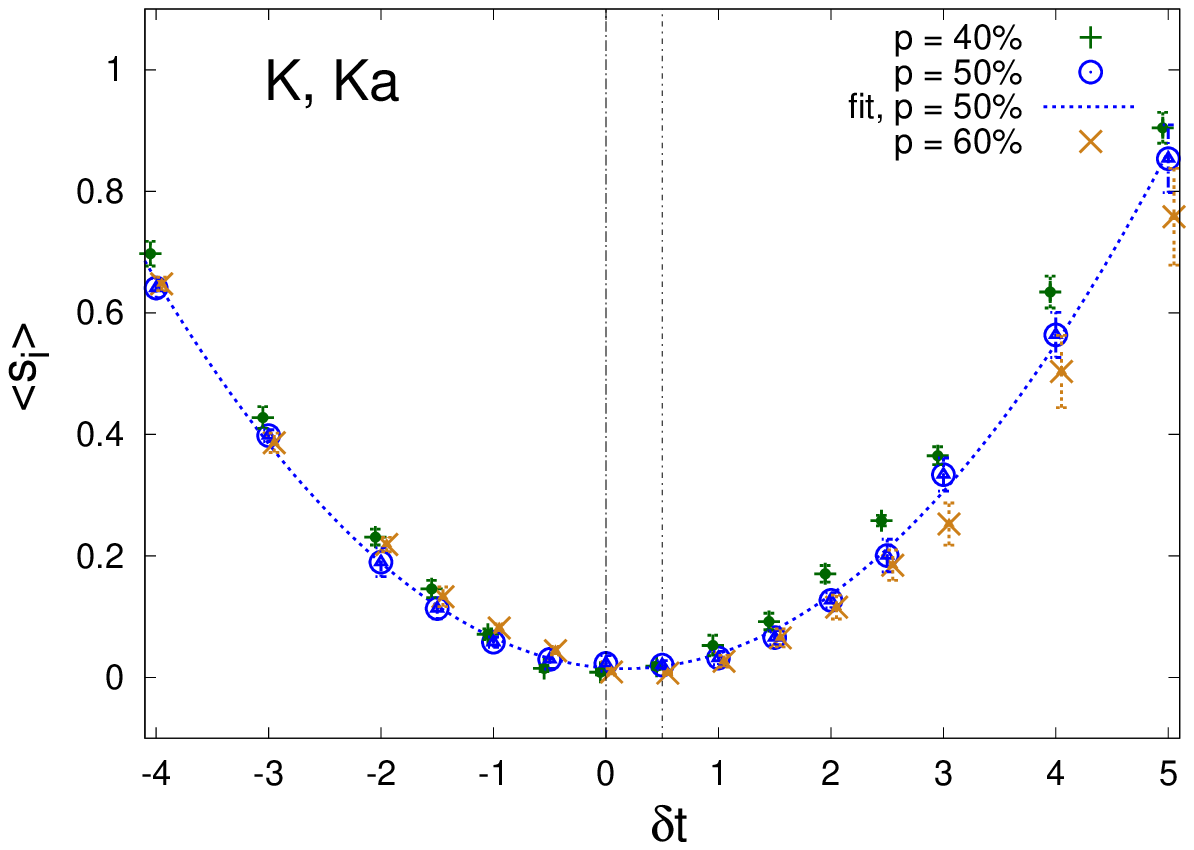}
\includegraphics[width=7cm]{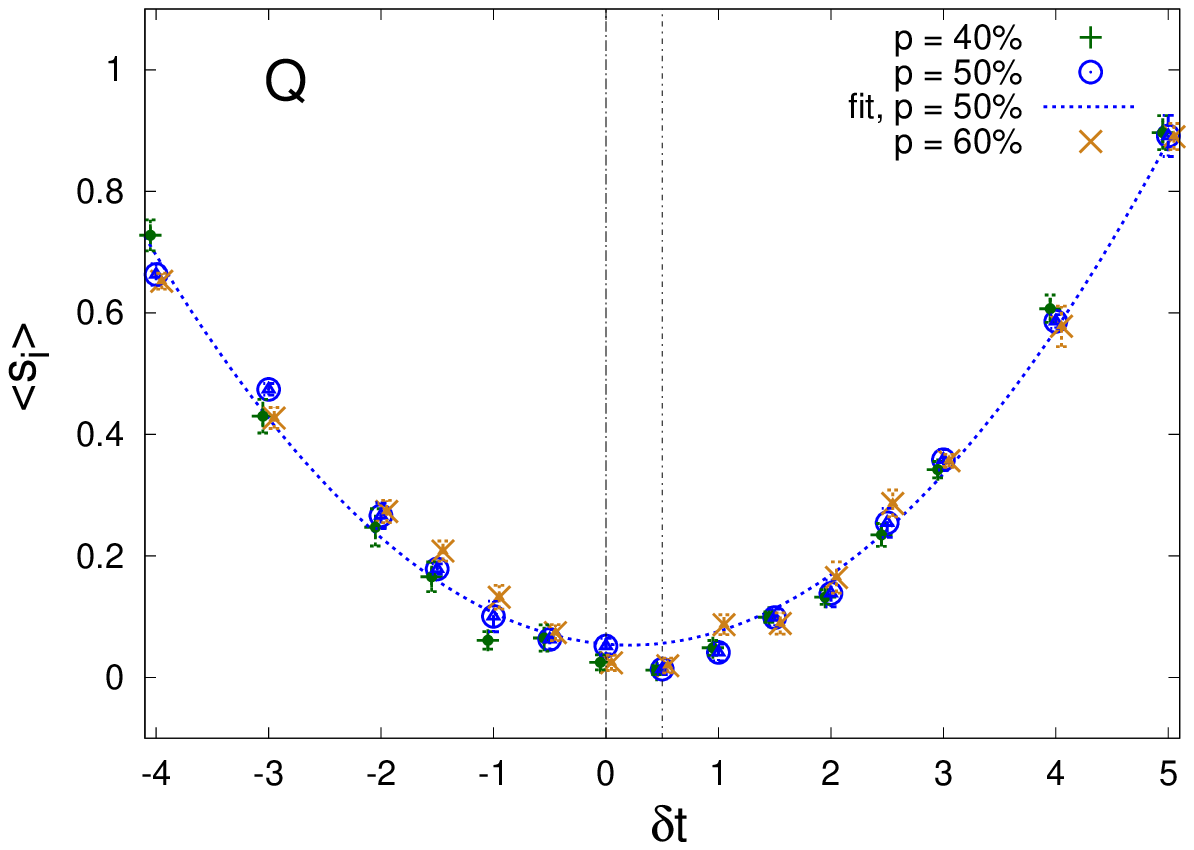}
\includegraphics[width=7cm]{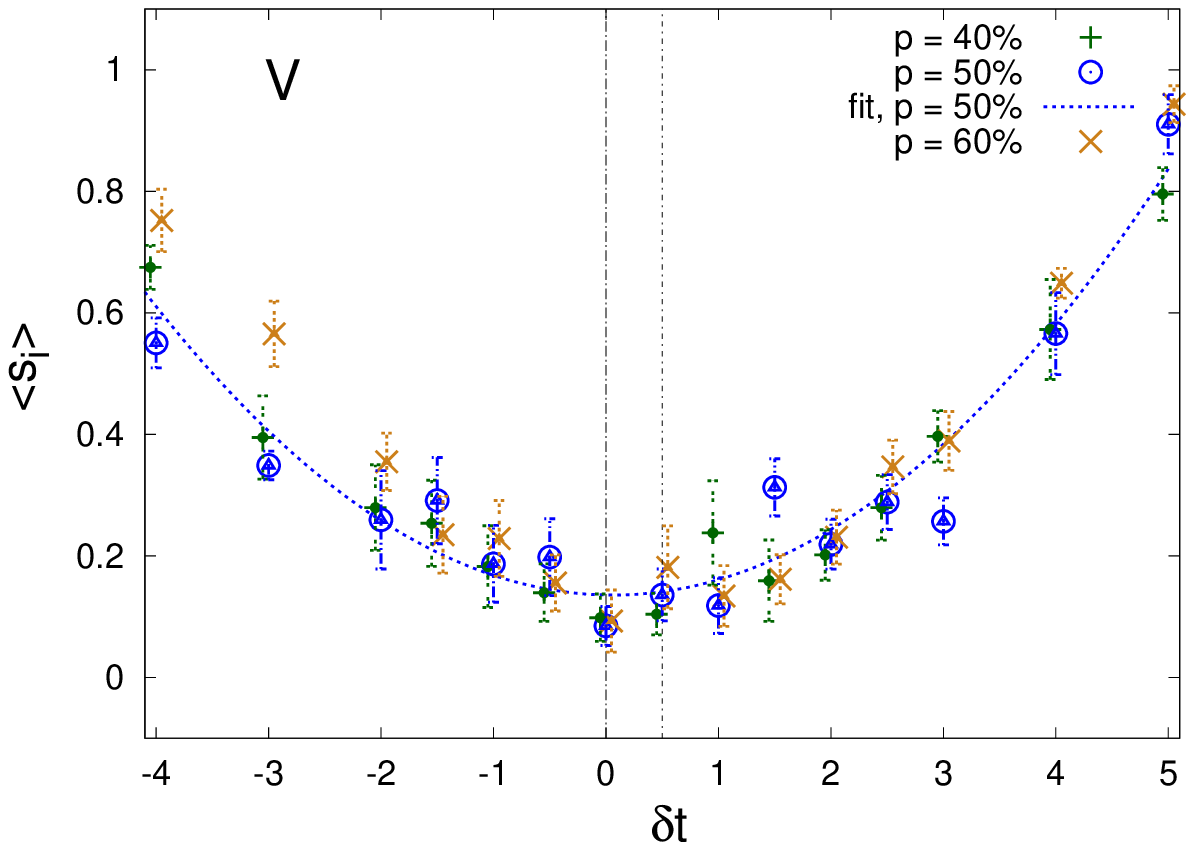}
\includegraphics[width=7cm]{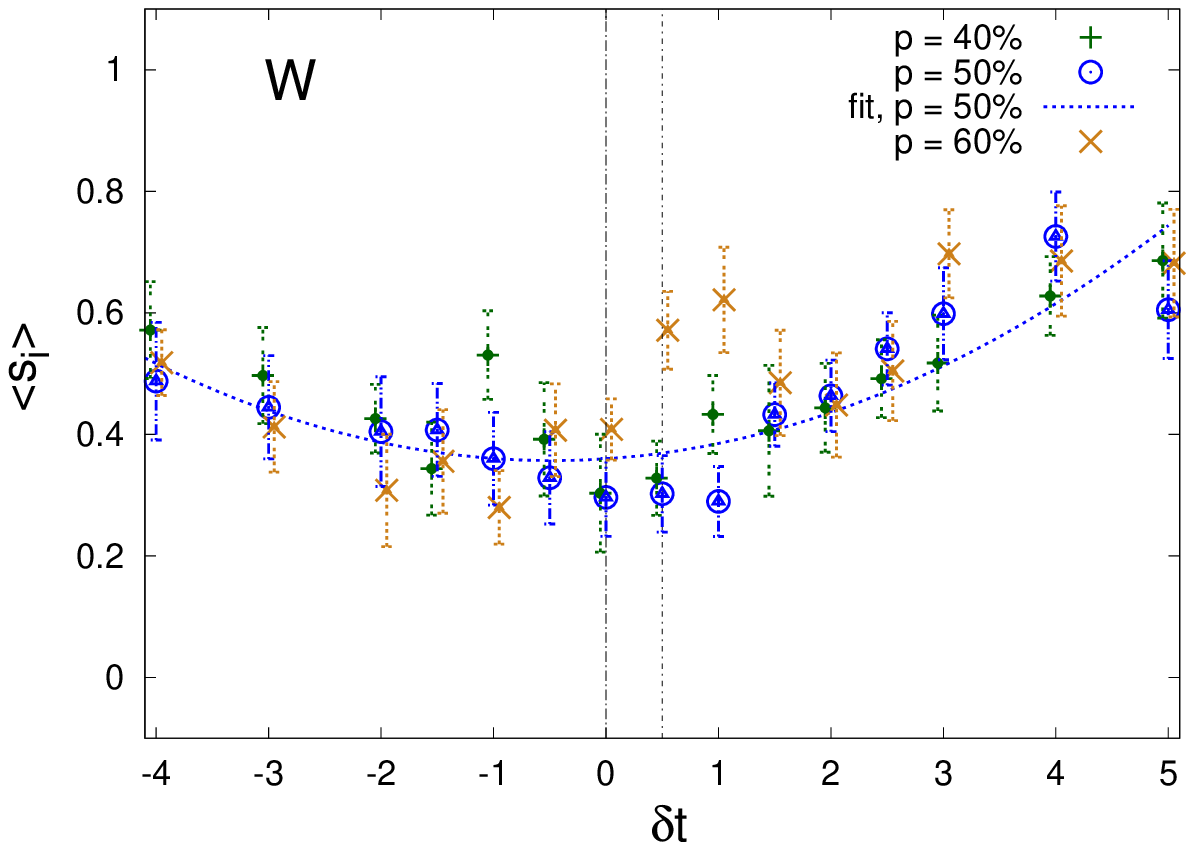}
\caption[]{ \mycaptionfont Mean and
  uncertainty of $s_i(\delta t)$ 
  [cf. Eq.~(\protect\ref{e-s-defn})]
  in the K and Ka, Q, V, and W
  bands, from top to bottom, respectively, as a function of $\delta t$
  in units of observing intervals. For each band, symbols show the
  mean and standard error in the mean for the 6 or 12 DA/year
  combinations in that band.  In addition to $s_i$ based on $\sigma_i$
  defined as the median standard deviation in a map ($\odot$, labelled
  ``$p= 50\%$''), equivalent statistics based on the 40\% ($+$) and
  60\% ($\times$) percentiles are also shown.  A least-squares
  best-fit parabola to the 50\% curve is shown in each panel
  {for convenience, but not used in estimating the minima $\tau_i$.}
  The
  values $\delta t=0$ and $\delta t=0.5$, favoured by
  \protect\nocite{LL10toffset}{Liu} {et~al.} (\protect\hyperlink{hypertarget:LL10toffset}{2010a}) [version (i)] and the WMAP collaboration,
  respectively, are shown as vertical lines.  Offsets in $\delta t$ of
  $\pm 0.05$ are applied to the $40\%$ and $60\%$ percentile
  statistics {to reduce symbol overlap}.  }
\label{f-KQVW}
\end{figure} 
} 

\newcommand\fhistall{
  \begin{figure}
    \centering 
    \includegraphics[width=8cm]{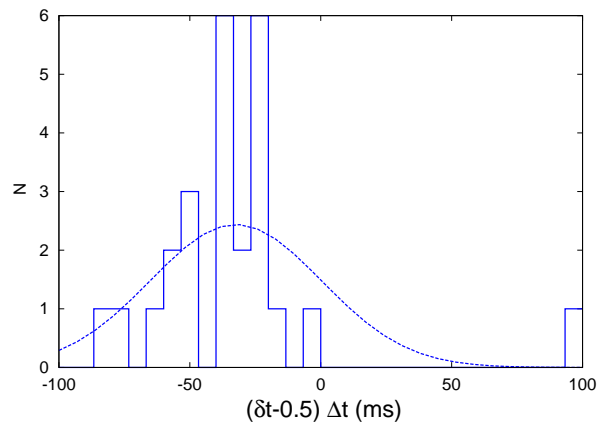}
    \caption[]{ \mycaptionfont \protect\prerefereechanges{Histogram of
        the timing offset estimates $\mu_{1/2}( \tau_i-0.5 ) \Delta t$
        from the individual DA/year combinations, of which per-band
        statistics are listed in Table~\protect\ref{t-KQVW}, and a
        Gaussian distribution appropriately normalised.}}
    \label{f-histall}
  \end{figure} 
} 

\newcommand\fWpmten{
  \begin{figure}
    \centering 
    \includegraphics[width=7cm]{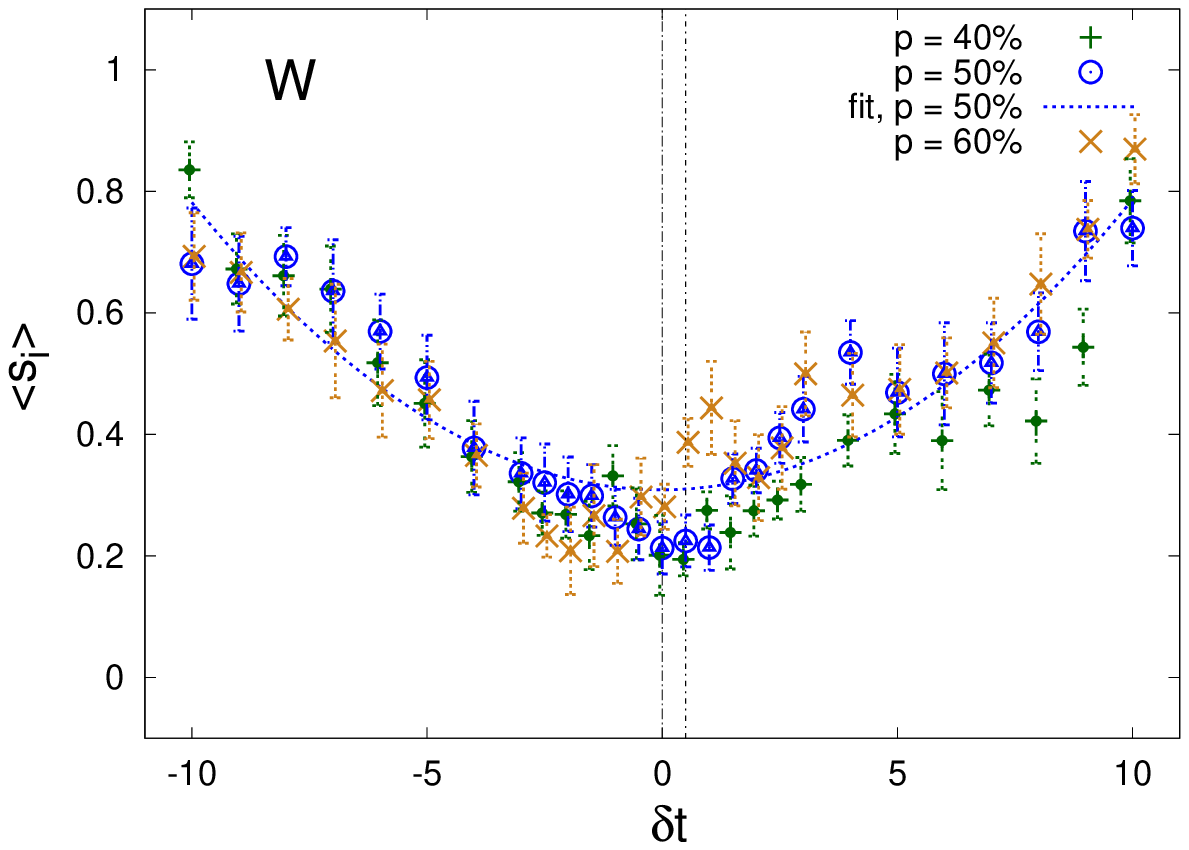}
    \caption[]{ \mycaptionfont Mean and uncertainty of
      $s_i(\delta t)$ 
      in the W band including {\em a posteriori} extra
      maps for large values of $|\delta t|$, as for Fig.~\protect\ref{f-KQVW}.
      \label{f-Wpmten}}
  \end{figure} 
} 

\newcommand\fvarmuKms{
\begin{figure}
  \centering 
  \includegraphics[width=8cm]{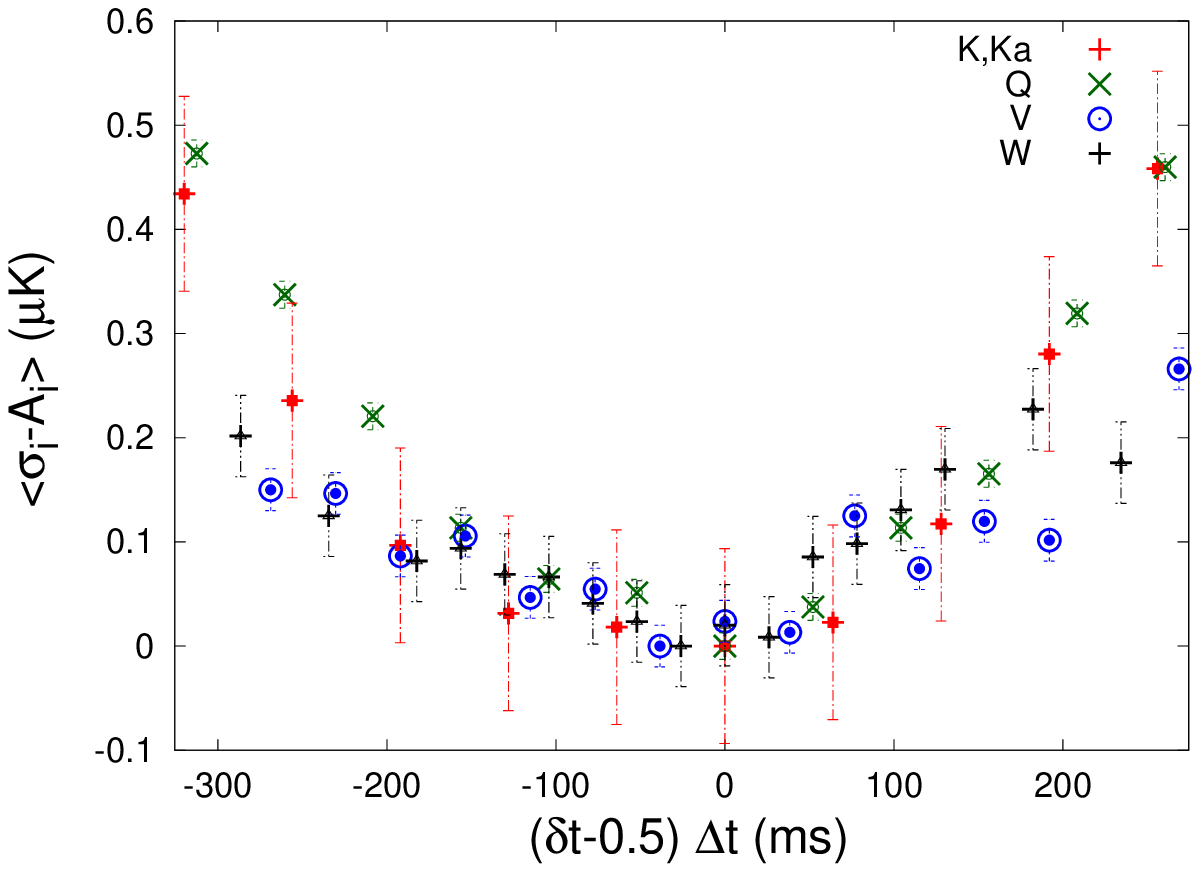}
  \caption[]{ \mycaptionfont Median standard deviation in $\mu$K with
    zeropoints removed, i.e. $\sigma_i(\delta t) -A_i$ as in
    Eq.~(\protect\ref{e-s-defn}), averaged (mean) over DA/year
    combinations in each of the K and Ka, Q, V, and W bands (symbols as
    indicated in the figure), as a function of timing offset in ms
    relative to the WMAP collaboration hypothesis, i.e. $(\delta
    t-0.5)\Delta t$. Error bars are standard errors in the mean. The
    timing offset range is centred on that of the hypothesis favoured
    from Table~\protect\ref{t-KQVW}, i.e.  $(\delta t-0.5)\Delta t =
    -25.6$.  }
  \label{f-varmuKms}
\end{figure} 
} 


\section{Introduction}  \label{s-intro}

\nocite{LL09lowquad}{Liu} \& {Li} (\protect\hyperlink{hypertarget:LL09lowquad}{2010a}) made sky maps out of the time-ordered data (TOD)
files of the Wilkinson Microwave Anisotropy Probe (WMAP) observations
of the cosmic microwave background (CMB) \nocite{WMAPbasic}({Bennett} {et~al.} \protect\hyperlink{hypertarget:WMAPbasic}{2003b}), following
the procedure recommended by the WMAP collaboration. They found a
considerably weaker CMB quadrupole signal than that estimated by the
WMAP collaboration, but were not aware of the reason for the
difference, given that their software pipeline satisfied many
tests. Later, \nocite{LL10toffset}{Liu} {et~al.} (\protect\hyperlink{hypertarget:LL10toffset}{2010a}) found that (i) they had used a
different choice of timing interpolation than that of the WMAP
collaboration, and (ii) there is a timing offset of $-25.6$~ms between
the spacecraft attitude and radio flux density timestamps,
recorded in the Meta and Science Data Tables, respectively, in the TOD
files.\footnote{The sign is chosen as the Science Data Table time minus
the Meta Data Table time.}
Individual observations in the Q, V and W bands last for 102.4~ms, 76.8~ms,
and 51.2~ms, respectively \nocite{WMAPExplSupp100405}(Section 3.2,
 {Limon} {et~al.} \protect\hyperlink{hypertarget:WMAPExplSupp100405}{2010}), so the offset corresponds to half
of a W band observing interval or a quarter of a Q band observing
interval. The 21 April 2010 version of the WMAP Explanatory Supplement 
\nocite{WMAPExplSupp100405}(Section 3.1,  {Limon} {et~al.} \protect\hyperlink{hypertarget:WMAPExplSupp100405}{2010}) 
does not refer to the $-25.6$~ms offset.

\nocite{LL10toffset}{Liu} {et~al.} (\protect\hyperlink{hypertarget:LL10toffset}{2010a}) presented a simple model for the effect of a
timing error on subtraction of the Doppler-induced
dipole signal, which is \prerefereechanges{celestial-position--dependent
because it is a dipole and} time-dependent because of the spacecraft's
orbit around the Sun, demonstrating that an artificial quadrupole-like
signal would be generated.
\nocite{MSS10}{Moss} {et~al.} (\protect\hyperlink{hypertarget:MSS10}{2010}) presented a toy model that confirmed
that a timing offset of $-25.6$~ms would induce an artificial quadrupole
approximately aligned with what had been previously considered to be
the cosmological CMB quadrupole signal. There was no clear
consensus between \nocite{LL10toffset}{Liu} {et~al.} (\protect\hyperlink{hypertarget:LL10toffset}{2010a}) and \nocite{MSS10}{Moss} {et~al.} (\protect\hyperlink{hypertarget:MSS10}{2010}) regarding the
fraction of the quadrupole that would be artificial.

If the timing offset induced an error in compiling the calibrated TOD
into maps, then this should introduce a blurring effect 
at a few-arcminute scale. This effect would not be easy to see by eye in sky maps.
However, statistically, in full-sky maps created without ignoring bright
objects, the hypothesis that a timing offset introduced an error in
the compilation of the calibrated TOD into maps was excluded to very
high significance \nocite{Rouk10Toffa}({Roukema} \protect\hyperlink{hypertarget:Rouk10Toffa}{2010}). 

Nevertheless, this method did not exclude the possibility that a timing
offset could have induced a quadrupole-like artefact during 
the {\em calibration} of
the uncalibrated TOD. Indeed, \nocite{Jarosik06Beam}{Jarosik} {et~al.}
\nocite{Jarosik06Beam}(Section 2.4.1, Figs~3, 4,  {Jarosik} {et~al.} \protect\hyperlink{hypertarget:Jarosik06Beam}{2007}) showed that an error
in the gain model could lead to a quadrupole-like artefact in inferred sky maps.
Although the Doppler dipole is not subtracted from the data during
the calibration step, it is the signal used for the calibration itself.

In principle, it would be possible to recalibrate the uncalibrated TOD
using several different values of the timing offset, remake sky maps
from the newly calibrated TOD, and test a statistic of these sky maps
that shows which timing offset is optimal. However, a simpler approach
is possible.

A timing error during calibration would not only induce an
artificial quadrupole-like signal in the mean. It would also add variance
to the sky map signal per pixel. This would be present in the calibrated TOD.
In other words, the effect of the error would be to add a small noise component
to the signal per pixel.
Since the calibrated TOD consist of individual observations, as do the uncalibrated TOD,
reversing an incorrect timing offset when map-making from the calibrated TOD
should approximately
correct the dominant signal, the dipole, reducing this variance. 
Here, minimalisation of the variance per pixel, as a function of 
timing offset, is carried out. Maps of both $\delta T$
and $(\delta T)^2$ are generated from the calibrated TOD as a
function of timing offset in order that variance maps can be
inferred. The timing offset that gives the minimum variance should be
that which is correct. 

The would-be timing-offset--induced calibration error and
the way that an approximate reversal of the error should reduce
the per-pixel variance are explained in \SSS\ref{s-method-principle}.
A toy model illustrating the variance reduction
is presented in
\SSS\ref{s-method-toy}. The observational analysis and 
variance minimisation method are described in 
\SSS\ref{s-method-minvar}. 
{The timing offset hypotheses are summarised
quantitatively in Table~\ref{t-hypothesis}.}
Results are presented in \SSS\ref{s-results}, 
{with the main results in 
Table~\ref{t-KQVW}.}
A rough estimate of the relevance of the results for the 
CMB quadrupole is made in \SSS\ref{s-disc-quad}.
\prerefereechanges{The discussion section deals with
sensitivity to the fitting method (\SSS\ref{s-disc-fit-sensitivity}),
consistency between wavebands (\SSS\ref{s-discuss-cf-wavebands}),
the possible relevance of sidelobe and other beam effects
(\SSS\ref{s-disc-sidelobe}), consistency with the results 
of using an alternative method (\SSS\ref{s-disc-manyiters}), 
and various caveats (\SSS\ref{s-disc-alternatives}).}
Conclusions are given in \SSS\ref{s-conclu}.
Gaussian error distributions are assumed throughout,
\prerefereechanges{except where otherwise stated.}

\section{Method} \label{s-method}

\subsection{Model of would-be calibration error} \label{s-method-principle}

The full calibration procedure is described \prerefereechanges{in 
\nocite{WMAP03syserr}{Hinshaw} {et~al.} (\protect\hyperlink{hypertarget:WMAP03syserr}{2003}), along with details for the 3-year data in 
\nocite{Jarosik06Beam}{Jarosik} {et~al.} (\protect\hyperlink{hypertarget:Jarosik06Beam}{2007}).}
Calibration to 
\prerefereechanges{the dipole signal} plays a 
fundamental role in this procedure. 
The direction and amplitude of this signal vary slightly throughout the year, 
mainly as a result of the orbit of the WMAP spacecraft around the 
\prerefereechanges{Solar System barycentre.}

\prerefereechanges{The dipole signal used in the calibration procedure
includes both an estimate of the contribution from the Solar System barycentre's
motion with respect to the comoving coordinate system, and a 
component due to the WMAP spacecraft's motion around the 
Solar System barycentre \nocite{WMAP03syserr}(Sect.~2.2.1 {Hinshaw} {et~al.} \protect\hyperlink{hypertarget:WMAP03syserr}{2003}).
The latter is measured to high precision, presumably as a function 
of a time standard, e.g. Julian days, consistent with that
used in the main mapmaking procedure.

Let us suppose that a timing offset is present when calculating the 
dipole that is used for the calibration of one hour of TOD, i.e. the
times associated with the 
TOD spacecraft attitude quaternions are used rather than the times
associated with the TOD.
Suppose that the only signal observed is that of the
true dipole, and that the random errors in measurement are negligible. 
Over the approximately annular region corresponding to one hour's observations,
the result of the calibration 
will be to shift the input signal in sky position to match
the dipole calculated for a slightly incorrect position. 
That is, the gain and baseline will be wrongly estimated
by a small amount that depends on the wrongly calculated dipole expected
to be present in the observed annulus, in a way that on average
corresponds to a positional shift.

Over a full year, the set of observations in a given 
pixel (for some given pixel size) depends on the various hourly 
subsets of TOD of which that pixel is a member. 
The set of temperature fluctuation estimates in that pixel
will have non-zero variance
induced by the varying errors from different hourly calibrations.
Thus, by calibrating with a timing offset, 
noise per pixel is added to observations that by assumption
consist of the true dipole and}
negligible random measurement error.

Now suppose that we have a set of wrongly calibrated observations
(TOD), retaining the assumptions that the only observed signal is the
dipole and that measurement error is negligible. Consider the meaning
of calibration. 
\prerefereechanges{In any hourly subset of TOD,}
the measured signal is offset
and scaled in such a way that it matches the calibration signal.
Thus, we know what the result of wrongly calibrating the input signal
is. It is the assumed dipole. Hence, the wrongly calibrated values
can be corrected 
\prerefereechanges{towards} the true values by using knowledge of the scanning
path, the assumed dipole, and the timing offset. This reduces the 
variance per pixel down 
\prerefereechanges{towards} the original, negligible measurement error.
\prerefereechanges{The error cannot be reduced to zero, in this simplified
case, because the calibration is not carried out on individual observations
(which would be meaningless), it is carried out on (usually) hourly 
subsets of the TOD.}

\prerefereechanges{This can be expressed mathematically as follows.  
Given errorless observations $\delta T[\phi(t), \theta(t)]$ of the dipole,
the hourly calibration of these correct dipole observations to 
an incorrectly timed
dipole is equivalent to applying the
functional $ \gh \approx g$, where 
$\gh$ is the functional for the hourly calibration, 
\begin{eqnarray}
  g : \delta T[ \phi(t), \theta(t) ] 
  \mapsto
  \delta T[ \phi(t - \delta t), \theta(t - \delta t) ] 
\label{e-shift-defn}
\end{eqnarray}
is the functional for what would hypothetically consist of 
calibrating every observation individually, 
and the timing offset is $\delta t$. This approximation 
should hold because the dipole is a smooth function with
features only on large scales.}
Thus, the wrongly calibrated TOD can be written as $\gh(\delta T)$. 

\ftoy

The inversion of \prerefereechanges{the shift $g^{-1}$ is
\begin{equation}
  g^{-1} : \delta T[ \phi(t), \theta(t) ] \mapsto 
  \delta T[ \phi(t + \delta t), \theta(t + \delta t) ] .
\label{e-shift-inv-defn}
\end{equation}
Thus, 
\begin{eqnarray}
  g^{-1} \{\gh [\delta T(\phi,\theta)]\} 
  & \approx & g^{-1} \{g [\delta T(\phi,\theta)]\} \nonumber \\
  & = & \delta T(\phi,\theta).
  \label{e-approximate-inversion}
\end{eqnarray}
Hence,
\begin{eqnarray}
  \sigma \left( g^{-1} \{\gh [\delta T(\phi,\theta)]\} \right) 
  \Big|_{\phi,\theta} 
  & \approx &
  \sigma \left[\delta T(\phi,\theta)\right]
  \Big|_{\phi,\theta} \nonumber \\
  & <  & \epsilon \nonumber \\
  & < & 
  \sigma \left\{ g [\delta T(\phi,\theta)] \right\}
  \Big|_{\phi,\theta} ,
  \label{e-minimisation-argument}
\end{eqnarray}
where} the variance $\sigma^2$ is defined for the set of observations
in a pixel at a fixed position $(\phi,\theta)$.  The relation $ \sigma
\left[\delta T(\phi,\theta)\right] \Big|_{\phi,\theta} < \epsilon $ is
the assumption that the observations have negligible ($ < \epsilon \ll
1$ in any units) measurement error.  Hence,
\prerefereechanges{given a set of wrongly calibrated
  observations $\delta T ' := \gh(\delta T)$ of originally errorless}
  observations $\delta T$ of the pure dipole signal, minimisation of
\begin{equation}
  \sigma \left[ g^{-1} (\delta T') \right] 
  \Big|_{\phi,\theta}
\end{equation}
as a function of $\delta t$ should yield the timing offset implicit 
in the wrong calibration. 

In reality, the signal is not expected to be a pure dipole signal, and
the measurements are not errorless. 
\prerefereechanges{Moreover, an iterative procedure involving both calibration
and mapmaking is used to optimise the calibration procedure
\nocite{WMAP03syserr}(Sects~2.2, 2.3,  {Hinshaw} {et~al.} \protect\hyperlink{hypertarget:WMAP03syserr}{2003}), and a more complex
gain model is used in the calibration of the 3-year data 
\nocite{Jarosik06Beam}(Sect~2.4,  {Jarosik} {et~al.} \protect\hyperlink{hypertarget:Jarosik06Beam}{2007}).
Thus, some additional error will be introduced into the
above minimisation by the complex calibration procedure.}

\prerefereechanges{Another potential 
complication is that use of $g^{-1}$ will also induce 
a blurring effect} at small (arcminute) scales
\nocite{Rouk10Toffa}({Roukema} \protect\hyperlink{hypertarget:Rouk10Toffa}{2010}, and references therein). 
At this small scale, application of $g^{-1}$ 
should increase the variance per pixel if the dipole is assumed to
be constant on this scale.
Hence, in order to detect the original calibration error, 
a large enough pixel size must be used so that the dominant contribution to
the variance per pixel is from positional offsets to the dipole rather than
from blurring of point sources.

\subsection{Toy model} \label{s-method-toy}

\prerefereestart
A toy model illustrating the detection of a timing offset by minimising
the variance per pixel
can be set up as
follows. 
Consider a fixed map vector $m$ of $\Npix$ pixels, 
composed of 
a fixed sinusoidal ``dipole'' map (a one-dimensional simplification of a
dipole) defined
\begin{equation}
\mapdipole(j) := 3 \sin\left( 2\pi \frac{j}{\Npix} \right),
  \label{e-fixed-dipole-map}
\end{equation}
at pixel $j$ 
and a component that includes CMB signal and noise, 
modelled (for simplicity)
as Gaussian noise of zero mean and standard deviation 0.5, i.e. 
$m$ is pixelwise defined
\begin{equation}
  m(j) := \mapdipole(j) + 0.5 G(0,1).
  \label{e-map-with-noise}
\end{equation}
The units can be thought of as mK.

A TOD vector $d$ representing
$\Ntod$ successive observations during one 
year can now be defined 
via an 
$\Ntod \times \Npix$ mapping matrix 
$M(\delta t)$ at timing offset $\delta t$
\begin{equation}
  d = M(\delta t)\, m + 
  \left[ 
    M(\delta t) \cdot {\mapannualdipole(\delta t)}^{\mathrm{T}} \right]
  \, \mathbf{e}
\end{equation}
where $\cdot$ is
the Hadamard product\footnote{Entry-wise product.},
\begin{equation}
  \left[\mapannualdipole(\delta t)\right]_{ij} := 0.3 \sin \left[
    2\pi \frac{j}{\Npix} + 2\pi \frac{i + \delta t}{\Ntod} 
    \right]
\end{equation}
is an $\Npix \times \Ntod$ matrix representing
the dipole component that varies during the year,
and $\mathbf{e}$ is a column vector of size $\Npix$ whose entries
are 1, used for summation.
The timing offset $\delta t$ may either be the correct value, 
in which case $M(\delta t)$ is the correct
mapping matrix
and $\mapannualdipole(\delta t) $ is the correct time-varying dipole,
or $\delta t$ may be a guessed value.

Let the mapping matrix $M(\delta t)$
be initialised to zero. For the $i$-th TOD observation, define the
first beam column number
\begin{eqnarray}
  && j_1(\delta t)  :=  \nonumber \\
  && \mathrm{mod} \left\{ \left\lfloor 
    \frac{70.5}{360}  \Npix 
    \left[
    \sin \left( 100 \pi \frac{i+\delta t}{\Ntod} \right) +  \frac{i}{\Ntod} 
    \right]
    \right\rfloor,
    \; \Npix \right\}, \nonumber \\
  \label{e-defn-j1}
\end{eqnarray}
where $\lfloor . \rfloor $ is the floor function,
set the second beam column number
\begin{equation}
  j_2[j_1(\delta t)] 
  := 
  \mathrm{mod} 
  \left[ j_1(\delta t) + \left\lfloor \frac{141}{360} \Npix \right\rfloor,
    \Npix \right]
  \label{e-defn-j2}
\end{equation}
and set matrix elements for the two beams
\begin{eqnarray}
  M_{i,j_1(\delta t)} &=& 1/(1 + \xim)    \nonumber \\
   M_{i,j_2[j_1(\delta t)]} &=& -1/(1-\xim),
  \label{e-defn-M}
\end{eqnarray}
where $\xim = \pm 0.007$ is a \prerefereechanges{differencing
assmbly} imbalance parameter with a randomly
chosen sign in a given simulation.  The motivation for these functions
is to have a scanning pattern whose fractional sky coverage and
fractional year coverage roughly correspond to those of the real
observations, e.g. one beam varies approximately sinusoidally in addition 
to a linear ($S^1$) yearly orbit, and the
second beam is separated from the first by a pixel distance
of $(141\ddeg/360\ddeg)\Npix$.  

The first and second moments of temperature
fluctuation estimates calculated directly from the calibrated TOD $d$
are maps $\muone$ and $\mutwo$, each of whose
$j$-th pixel is
\begin{eqnarray}
  && \muone(\delta t)  :=  \nonumber \\
  && \frac{\sum_i 
    \left(M^{\mathrm{T}}(\delta t) 
    \left\{d -M(\delta t) \,\mapdipole
      -  
      \left[ M(\delta t) \cdot {\mapannualdipole(\delta t)}^{\mathrm{T}} \right]
      \, \mathbf{e}
      \right\} \right)_{ij} 
  }{\sum_i 1} \nonumber \\
  \label{e-1moment-defn}
\end{eqnarray}
and
\begin{eqnarray}
&&  \mutwo(\delta t)  :=   \nonumber \\
&& \rule{-3ex}{0ex} \frac{\sum_i 
    \left[ 
    \left(M^{\mathrm{T}}(\delta t) 
    \left\{d -M(\delta t) \,\mapdipole
      -  \left[ M(\delta t) \cdot {\mapannualdipole(\delta t)}^{\mathrm{T}} \right]
      \, \mathbf{e}
      \right\} \right)_{ij} 
    \right]^2 
  }{\sum_i 1},  \nonumber \\
  \label{e-2moment-defn}
\end{eqnarray}
respectively, where the sums are taken over the rows $i$ where
\begin{eqnarray}
  \left(M^{\mathrm{T}}(\delta t) 
  \left\{d -M(\delta t) \,\mapdipole
  -  \left[ M(\delta t) \cdot {\mapannualdipole(\delta t)}^{\mathrm{T}} \right]
  \, \mathbf{e}
  \right\} \right)_{ij} 
  \not= 0, \nonumber \\
\end{eqnarray} 
and $(\phi,\theta)$ is the celestial position of pixel $j$.
Given the TOD $d$ for an unknown
timing offset, and the dependence of the mapping matrix on timing
offset $M(\delta t)$ from Eqs~(\ref{e-defn-j1}), (\ref{e-defn-j2}),
and (\ref{e-defn-M}), the standard deviation map can be estimated
element-wise
\begin{equation}
  \sigperpix := \sqrt{ \left[\mutwo(\delta t)\right]_j - \left[\muone(\delta t)\right]_j^2 }
  \label{e-sigperpix}
\end{equation}

For each $\delta t$, define the median per map of the
standard deviation per pixel $\sigperpix$
\begin{equation}
\sigma(\delta t) : = \mu_{1/2} \left( \sigperpix \right).
\label{e-defn-sigma-median}
\end{equation}
Fitting a simple symmetrical function, i.e. a parabola, 
to  $\sigma(\delta t)$,
should yield an estimate of
$\delta t$ that minimises $\sigma(\delta t)$. 
This can be compared to the known value input to the toy model.
Figure~\ref{f-toy} shows the dependence of 
 $\sigma(\delta t)$
on $\delta t$
for a realisation of the
model defined here, with $\Npix=20, \Ntod=2000$, 
and an input timing offset $\delta t = 0.5$. 
The range of timing offsets
is chosen to be symmetrical around $\delta t = 0.5$.
There is clearly a minimum in $\sigma$ close to $\delta t = 0.5$. 
Over 30 simulations with these same parameters, 
the median and standard error in the median of the 
$\delta t$ values that minimise the best-fit parabola to 
the median per map of 
$\sigma(\delta t)$ are 
$0.500\pm 0.003$.    

\prerefereestop

\thypothesis

\subsection{Observational data and variance minimisation} \label{s-method-minvar}
As in \nocite{Rouk10Toffa}{Roukema} (\protect\hyperlink{hypertarget:Rouk10Toffa}{2010}), only the three-year WMAP TOD are analysed, in order to
reduce the computing load. However, the full-year,
filtered, calibrated
TOD\footnote{\href{http://lambda.gsfc.nasa.gov/product/map/dr2/tod_fcal_get.cfm}{{\tt http://lambda.gsfc.nasa.gov/product/map/dr2/}}
\href{http://lambda.gsfc.nasa.gov/product/map/dr2/tod_fcal_get.cfm}{{\tt tod\_fcal\_get.cfm}}}
are analysed (not just 198 or 199 days),
for all three years.  Analyses are made using the K, Ka, Q, V and W bands. 
Since there is only one
differencing assembly (DA) for each of the K and Ka wavebands, giving only three 
one-year data sets each, these are combined together to form a single set of six DA/year 
sets.\footnote{For brevity, the K and Ka bands will sometimes be referred to below as a single band.}
The Q and V bands each have two DAs, giving six DA/year sets each, and
the W band has four DAs, giving twelve DA/year sets.

For any DA/year set, for a range of timing offsets, the filtered, calibrated TOD 
are compiled using a patch to 
\nocite{LL10toffset}{Liu} {et~al.} (\protect\hyperlink{hypertarget:LL10toffset}{2010a})'s 
publicly available data analysis pipeline.\footnote{\url{http://cosmocoffee.info/viewtopic.php?p=4525}, 
\href{http://dpc.aire.org.cn/data/wmap/09072731/release_v1/source_code/v1/}{{\tt http://dpc.aire.org.cn/data/wmap/09072731/release\_v1/}}
\href{http://dpc.aire.org.cn/data/wmap/09072731/release_v1/source_code/v1/}{{\tt source\_code/v1/}} and associated libraries}
The patch is optimised for the present analysis and enables execution of the pipeline
using the GNU Data Language 
(GDL).\footnote{\href{http://cosmo.torun.pl/GPLdownload/LLmapmaking_GDLpatches/LLmapmaking_GDLpatches_0.0.4.tbz}{{\tt http://cosmo.torun.pl/GPLdownload/}}
\href{http://cosmo.torun.pl/GPLdownload/LLmapmaking_GDLpatches/LLmapmaking_GDLpatches_0.0.4.tbz}{{\tt LLmapmaking\_GDLpatches/LLmapmaking\_GDLpatches\_0.0.4.tbz}}}
In contrast to the analysis in \nocite{Rouk10Toffa}{Roukema} (\protect\hyperlink{hypertarget:Rouk10Toffa}{2010}), standard masking
is retained in order to avoid planet observations ({\sc daf\_mask})
and the Galactic
Plane.\footnote{\protect\href{http://lambda.gsfc.nasa.gov/data/map/dr2/ancillary/wmap_processing_r9_mask_3yr_v2.fits}{{\tt
      http://lambda.gsfc.nasa.gov/data/map/dr2/}}
  \protect\href{http://lambda.gsfc.nasa.gov/data/map/dr2/ancillary/wmap_processing_r9_mask_3yr_v2.fits}{{\tt
      ancillary/wmap\_processing\_r9\_mask\_3yr\_v2.fits}}} The
``pessimistic'' mode of downgrading the processing mask resolution is
used.  \prerefereechanges{In order to calculate variances, each
  mapmaking step creates both a mean signal map and a mean square
  signal map, as in 
  Eqs~(\ref{e-1moment-defn}) and (\ref{e-2moment-defn}).
  This is equivalent to using the first map iteration. 
  
  Use of a highly-iterated map estimate would be equivalent to adding
  an extra term that would vary with $\delta t$ in a complex way, to
  these equations.  This would add an additional source of noise to
  the statistic, decreasing its statistical power. Nevertheless, given
  that the calibration is made using a complex iterative process both
  of sky maps and baseline and gain parameters, it should be useful to
  see if this gives a result that is statistically compatible with the
  more accurate results using Eqs~(\ref{e-1moment-defn}) and
  (\ref{e-2moment-defn}), i.e. without addition of the extra
  term. Thus, an extra set of calculations has been made using the
  80-th mapmaking iteration for the bands with the smaller numbers of
  DA's and observations. }

\fKQVW

\nocite{LL10toffset}{Liu} {et~al.} (\protect\hyperlink{hypertarget:LL10toffset}{2010a})'s convention for labelling the timing offset,
including the sign, is retained. 
That is, the {\sc center} parameter for determining the timing offset
is generalised to an arbitrary floating point value,
written here as $\delta t$, expressed as a fraction of 
an observing time interval in a given band. 
\nocite{LL10toffset}{Liu} {et~al.} (\protect\hyperlink{hypertarget:LL10toffset}{2010a})'s hypothesis of what is the correct offset, interpreted
to mean that the true error occurred during the calibration step and not during the 
mapmaking step, includes two versions, labelled (i) and (ii) above
(\SSS\ref{s-intro}).
As (i) a ``timing interpolation'', the hypothesis 
can be defined to mean $\delta t = 0$ in all bands.
The WMAP collaboration's preferred timing offset 
corresponds to $\delta t = 0.5$.
The second version of \nocite{LL10toffset}{Liu} {et~al.} (\protect\hyperlink{hypertarget:LL10toffset}{2010a})'s hypothesis, (ii) above
(\SSS\ref{s-intro}), is the TOD file 
fixed timing offset relative to the WMAP
collaboration's choice, i.e. 
$(\delta t -0.5)\Delta t = -25.6$~ms,
where $\Delta t$ is the exposure time of a single observation in the K
and Ka, Q, V, or W band, i.e. 128~ms, 102.4~ms, 76.8~ms, or 51.2~ms,
respectively.  Thus, in the W band, 
$-25.6$~ms relative to $\delta t=0.5$
corresponds to $\delta t = 0$, while at lower frequencies, the offset fixed 
in milliseconds
corresponds to values of $\delta t$ between 0 and 0.5.
{The different hypotheses of the timing offset are 
summarised in Table~\ref{t-hypothesis} as a function of waveband.}

As explained in \SSS\ref{s-method-principle}, the pixel sizes must be large
enough to avoid introducing variance due to the incorrect inversion of
small-scale signals (e.g. objects at the few-arcminute
scale are blurred by $g^{-1}$). 
Hence, a
HEALPix \nocite{Healpix98}({G\'orski} {et~al.} \protect\hyperlink{hypertarget:Healpix98}{1999}) resolution of $N_{\mathrm{side}}=8$, i.e.
pixel sizes of about $7.3\ddeg$, is adopted here. 
This should reduce the chance that small-scale effects contaminate the
effect of interest here.

For each DA/year combination $i$ in each band, maps of the signal 
$\muone$ 
and the square signal 
$\mutwo$ 
are calculated over 
\prerefereechanges{$-4 \le \delta t \le 5$, i.e. at values symmetric
around $\delta t = 0.5$, with smaller intervals closer to that
value.}
This wide range in $\delta t$ is used because, as for the blurring effect analysed
in \nocite{Rouk10Toffa}{Roukema} (\protect\hyperlink{hypertarget:Rouk10Toffa}{2010}), exaggerating the absolute timing offset should
strengthen the effect of using a wrong value.

For a given $\delta t$, the two maps are used to infer a map 
of the standard deviation per pixel 
$\sigperpix$ 
where $\delta T$ now represents the calibrated (whether correctly or not) TOD.
Let $\sigma_i(\delta t)$ be the median of this quantity over the map,
\prerefereechanges{as in Eq.~(\ref{e-defn-sigma-median}),}
where invalid pixels (mostly in the Galactic Plane) are ignored.

These ``median standard deviations''\footnote{The standard deviation is
calculated at a fixed position $(\phi,\theta)$. The median of this quantity
is calculated over all valid positions $(\phi,\theta)$.}
$\sigma_i(\delta t)$ 
are normalised over $\delta t$ internally within each sample $i$ 
in order to give approximately equal weight to the different
DA/year samples $i$ in a given band. That is,
the minimum $A_i$ and maximum $B_i$ of 
$\sigma_i(\delta t)$ are used to calculate
\begin{equation}
  s_i(\delta t) := \frac{ \sigma_i (\delta t) - A_i}{B_i-A_i}.
\label{e-s-defn}
\end{equation}

In order to estimate the minimum of $s_i(\delta t)$ in a given DA/year
combination in a given band, a smooth, symmetric function is needed.
{The simplest obvious choice is a parabola.}
Since $s_i$ is normalised to the
range $[0,1]$, the parabola $a_i (\delta t)^2 + b_i\delta t +c_i$ that
least-squares best-fits {$s_i(\delta t)$} is found, giving an
estimate 
\begin{equation}
\tau_i := -b_i/(2a_i)
\label{e-defn-tau}
\end{equation}
 of the $\delta t$ value that minimises $s_i(\delta t)$.
For a given band, each of the 
6 (K and Ka, Q, or V) or 12 (W) DA/year
combinations $i$ yields an estimate $\tau_i$. Under the assumption
of statistical independence among the DA/year combinations, the median $\mu_{1/2}( \tau_i )$
and standard error in the median $\sigma_{\mu_{1/2}( \tau_i )}$\footnote{Gaussian error distributions
are assumed, i.e. the standard error in the median is estimated as 1.253 times
the standard error in the mean.}
give an estimate of the optimal timing offset for the 
inversion of a would-be timing offset during calibration of
the uncalibrated TOD in that waveband.

\prerefereechanges{For the purposes of testing the sensitivity to the
fitting method (\SSS\ref{s-disc-fit-sensitivity}),
additional calculations at $\delta t = -5, -2.5$ were made 
in order to have a sample that is symmetric around $\delta t =0$, 
i.e. to test sampling sensitivity, for all wavebands.
Extra maps were also calculated
for $\delta t = -10, \ldots, -6, 6, \ldots, 10$ in W to see if a wider 
range in $\delta t$ gives a better defined minimum
(\SSS\ref{s-discuss-cf-wavebands}). 
}

\tKQVW

\fhistall

\section{Results} \label{s-results}

Calculations on 4-core, 2.4~GHz, 64-bit processors with 4~Gib RAM,
using GDL-0.9$\sim$rc4 running under GNU/Linux, took about 2--6 hours
per map, depending on waveband.\footnote{{The maps of the main
calculation and extra maps of the {\em a posteriori} extra W calculations 
(\SSS\protect\ref{s-discuss-cf-wavebands}) are available at
\href{http://cosmo.torun.pl/GPLdownload/LLmapmaking_GDLpatches/LXLoffset2_skymaps.tbz}{{\tt http://cosmo.torun.pl/GPLdownload/}}
\href{http://cosmo.torun.pl/GPLdownload/LLmapmaking_GDLpatches/LXLoffset2_skymaps.tbz}{{\tt LLmapmaking\_GDLpatches/LXLoffset2\_skymaps.tbz}}}.}
Figure~\ref{f-KQVW} clearly shows that the
per pixel variance of the maps is minimised near the preferred time
offsets of the WMAP collaboration and \nocite{LL10toffset}{Liu} {et~al.} (\protect\hyperlink{hypertarget:LL10toffset}{2010a}). Time
offsets of several observing intervals above or below these values
clearly increase the per pixel variance. \prerefereechanges{It is not clear in the
figure whether any of the claimed offset hypotheses is
significantly preferred as a sharp minimum. However, the approximate symmetry
axes of the function shapes clearly lie at lower $\delta t$ than
the central value $\delta t =0.5$.}

\fvarmuKms

 Table~\ref{t-KQVW} shows the statistics of $\tau_i$,
the minimum of $s_i$ for the individual DA/year combination for each
waveband.  Comparison with Table~\ref{t-hypothesis} shows that the
WMAP collaboration's preferred timing offset is rejected to high
significance in the lower frequency bands,
\prerefereechanges{i.e.  4.5$\sigma$ (K, Ka),
$8.2\sigma$ (Q), and $5.2\sigma$ (V), and not constrained in the W
band.  
Moreover, the K and Ka bands, and the Q band, also reject 
version (i) of \nocite{LL10toffset}{Liu} {et~al.} (\protect\hyperlink{hypertarget:LL10toffset}{2010a})'s hypothesis
to high significance, i.e. $3.0\sigma$ (K, Ka) and $7.6\sigma$ (Q).}
Hence, both the
WMAP collaboration's preferred timing offset and 
version (i) of \nocite{LL10toffset}{Liu} {et~al.} (\protect\hyperlink{hypertarget:LL10toffset}{2010a})'s
hypothesis
are rejected to high significance by minimisation of per pixel
variance of maps made from the calibrated, filtered TOD.

In contrast, 
the final {two} numerical columns in Table~\ref{t-KQVW} 
show that version (ii) of 
\nocite{LL10toffset}{Liu} {et~al.} (\protect\hyperlink{hypertarget:LL10toffset}{2010a})'s timing offset, defined as
the TOD file timing offset of $-25.6$~ms, is consistent with all
wavebands, 
\prerefereechanges{at 
$1.5\sigma, 0.5\sigma, 2.1\sigma,$ and $0.1\sigma$ in 
K and Ka, Q, V, and W, respectively.
The weighted mean
$\mu_{1/2}( \tau_i-0.5 ) \Delta t = -29.7 \pm 2.5$~ms rejects
the WMAP collaboration's hypothesis at $11.9\sigma$ and agrees with 
the fixed TOD file timing offset hypothesis at $1.6\sigma$.}

\prerefereechanges{The 
above significance estimates assume that the statistic $\tau$ 
is normally distributed. 
Figure~\ref{f-histall} shows the 25 timing offset estimates,
from individual 30 DA/year combinations,
that lie in the range $(-100,100)$~ms. The distribution is 
not well-modelled by a Gaussian distribution. A Kolmogorov-Smirnov
test with the mean and standard deviation of the data rejects the
hypothesis that the distribution is Gaussian with $p=1.4\%$. 
Could this affect the significance in rejecting the $\delta t =0.5$
hypothesis?

It is clear that it would be difficult to reconcile
the distribution in Fig.~\ref{f-histall} with having 
a median of 0~ms, especially 
given that the distribution is sampled many times. 
This can be tested more formally.
To non-parametrically test the hypothesis that the median of the
distribution is 0~ms, $10^6$ bootstrap 
resamples from the distribution were made. 
These give a 99.999\%
confidence interval for the median of the distribution to be
$\mu_{1/2}( \tau_i-0.5 ) \Delta t \in (-54,-20)$~ms.  
Thus, the WMAP $\delta t = 0.5$ hypothesis
can be conservatively rejected at $99.999\%$ confidence.
The same non-parametric method gives the $(2.5\%, 97.5\%)$ confidence interval
to be $(-47, -26)$~ms, i.e. 
taking into account \nocite{Bennett03MAP}{Bennett} {et~al.} (\protect\hyperlink{hypertarget:Bennett03MAP}{2003a})'s ({\protect\SSS}{6.1.2}) estimate of the
uncertainty in the timing as 1.7~ms, 
the hypothesis of a $-25.6$~ms offset is 
consistent with the data.}

\subsection{Quadrupole overestimation induced by the calibration error}
\label{s-disc-quad}

Unfortunately, the simple procedure presented here is only sufficient
to provide evidence for the timing-offset--induced calibration error,
not for compiling improved maps directly from the calibrated TOD.  As
explained in \SSS\ref{s-method-principle}, a timing-offset--induced
calibration error on a signal dominated by the dipole can be
approximately reversed, since the calibration is itself based on a
model dipole that is a good approximation to the measured dipole. 
However, the
remaining signal is not a dipole, so it is more difficult to correct
the remaining components of
the wrongly-calibrated signal.
As shown in \nocite{Rouk10Toffa}{Roukema} (\protect\hyperlink{hypertarget:Rouk10Toffa}{2010}), the present method of reversing
the calibration error, i.e. the use of $g^{-1}$, introduces a blurring
effect on small ($\sim 4\arcm$) scales.
To obtain CMB sky maps with the
calibration error fully removed, it is not obvious that there is a
simpler alternative to redoing the full calibration of the
uncalibrated TOD. 

Nevertheless, an {\em approximate} reversal of the $-25.6$~ms
timing-offset--induced calibration error in at least the dipole
component of the signal is possible using the same data analysis
pipeline as above. In this case, by how much does the quadrupole
decrease when the artificial quadrupole-like signal is approximately removed? 

Using the presently available version of the
WMAP 3-year, calibrated, filtered TOD in the W band, i.e. the band where
the cosmological signal is the least affected by foregrounds, 
the quadrupole can be calculated consistently for 
{maps at} both
$\delta t =0.5$, the incorrect timing offset, 
and $\delta t =0$, the timing offset
that yields approximately corrected maps.
{Using
maps made at a resolution of $N_{\mathrm{side}}=512$,}
\prerefereechanges{and subtracting the 3-year maximum entropy
synchrotron, free-free and dust maps provided by the WMAP
collaboration\footnote{\href{http://lambda.gsfc.nasa.gov/product/map/dr2/mem_maps_get.cfm}{{\tt http://lambda.gsfc.nasa.gov/product/map/dr2/}}
\href{http://lambda.gsfc.nasa.gov/product/map/dr2/mem_maps_get.cfm}{{\tt mem\_maps\_get.cfm}}},}
pseudo-$C_l$ estimates 
(after monopole and dipole removal)\footnote{The {\sc anafast} routine
of the GPL version of the {\sc HEALPix} software \nocite{Healpix98}({G\'orski} {et~al.} \protect\hyperlink{hypertarget:Healpix98}{1999}) was used here.} 
from the 3 years, for the 4 W band DA's,
are
\prerefereechanges{$(3/\pi) C_2 = 170 \pm 4 \,\mu$K$^2$ and $ 103 \pm 4 \,\mu$K$^2$, 
for $\delta t=0.5$ and $\delta t=0$, respectively, 
for the KQ85 sky mask, and
$(3/\pi) C_2 = 79 \pm 3 \,\mu$K$^2$ and $41 \pm 2 \,\mu$K$^2$} for 
the KQ75 sky mask,
where errors are standard errors in the mean over the 12 DA/year
combinations, and 80 map iterations are made for each DA/year
combination.  The dipole removal has little effect on these estimates.
Removal of only the monopole gives 
\prerefereechanges{
$(3/\pi) C_2 = 170 \pm 4 \,\mu$K$^2$ and $103 \pm 4 \,\mu$K$^2$, 
for $\delta t=0.5$ and $\delta t=0$, respectively, for the KQ85 sky mask, and
$(3/\pi) C_2 = 78 \pm 3 \,\mu$K$^2$ and $40 \pm 2\,\mu$K$^2$}
for the KQ75 sky mask.  Hence, the $-25.6$~ms
timing-offset--induced calibration error implies that estimates of the
CMB quadrupole amplitude $(3/\pi)C_2$ based on WMAP 3-year W band sky maps
derived from the incorrectly calibrated TOD are overestimated by
\prerefereechanges{$64\pm6\%$ and $94\pm10\%$}
for the KQ85 and KQ75 sky masks, respectively
(for removal of both the monopole and dipole).  

This is a significant and substantial systematic error. Given the
differences in method and data subsets [e.g., \nocite{LL09lowquad}{Liu} \& {Li} (\protect\hyperlink{hypertarget:LL09lowquad}{2010a})
  estimate cross-quadrupoles of the V and W five-year data, while the
  present estimate is the W 3-year auto-quadrupole], the KQ75 drop
in quadrupole power appears to be roughly consistent with that of {\SSS}{5.3} of
\nocite{LL09lowquad}{Liu} \& {Li} (\protect\hyperlink{hypertarget:LL09lowquad}{2010a}), i.e. greater than that of \nocite{MSS10}{Moss} {et~al.} (\protect\hyperlink{hypertarget:MSS10}{2010})'s toy
model estimate (Fig.~1 caption).  Moreover, the problem of large-scale
CMB power being concentrated towards the Galactic Plane is exacerbated
with this approximate correction of the calibration error. The
Galactic Plane is the part of the sky where difficult-to-estimate
systematic error can most be suspected.




\tdomainshift

\section{Discussion} \label{s-disc}

\subsection{{Sensitivity to fitting method}}
\label{s-disc-fit-sensitivity}

These results do not appear to be sensitive to the choice of symmetrical
fitting function. Least-squares hyperbolic best fits, i.e. best-fits of 
$a_i (\delta t)^2 + b_i\delta t +c_i$ to
$[s_i(\delta t)+0.5]^2$, where the vertical offset of 0.5 and positive square root 
of the best fit
give a top quadrant hyperbola, 
yield similar values to those in Table~\ref{t-KQVW}. For example,
over all wavebands, \prerefereechanges{this} hyperbolic fit gives
$\mu_{1/2}( \tau_i ) = 0.22 \pm 0.04$ and
$\mu_{1/2}( \tau_i-0.5 ) \Delta t = -29.4 \pm 4.3$~ms. 

\prerefereechanges{Since the $\delta t$ values used in sampling 
are symmetric 
around $\delta t =0.5$, it is difficult to see how this could induce a bias
against the $\delta t = 0.5$ hypothesis. Could it lead to an inaccurate estimate
in the case that 
$( \tau_i-0.5 ) \Delta t = -25.6$~ms is correct?
Table~\ref{t-domainshift} shows that using a $\delta t$ sample symmetric
around $\delta t =0$ gives results compatible with those 
in Table~\ref{t-KQVW}.}

Could the dependence of $s_i$ on $\delta t$ be asymmetric?
\prerefereechanges{The toy model (\SSS\ref{s-method-toy}) does not suggest
any obvious asymmetry, and it is not obvious what detailed differences between
the toy model and the real data could be sufficiently asymmetric to perturb
the method. Moreover,
the WMAP scan paths are curved (non-geodesic) paths on the
2-sphere
that have many symmetries. 
Nevertheless,} it is not
obvious that $s_i$ should vary with an exact symmetry around
its minimum, even if approximate symmetry seems reasonable. 
If the dependence of $s_i$ on $\delta t$ approached linearity for 
large $|\delta t|$, then one way of testing this could be to make
linear fits to subsets of data for large $|\delta t|$. However, 
comparison of parabolic and hyperbolic fits favours the former, i.e.
$s_i(\delta t)$ does not appear to approach linearity for large $|\delta t|$.
Thus, linear fits for large $|\delta t|$ would depend strongly on the 
$\delta t$ domains chosen. If an asymmetric model for $s_i$ dependence
on $\delta t$ could be found that is close enough to parabolic in shape
in order to provide fits in K and Ka, Q, and V as good as those 
shown in Fig.~\ref{f-KQVW}, then it could conceivably be possible
that the $\delta t = 0.5$ hypothesis could be made consistent
with the resulting minima. However, in that case, 
the fact that the symmetry assumption gives a solution
consistent with the timing offset recorded in the TOD files would
have to be attributed to coincidence.

\fWpmten

\subsection{Consistency between wavebands} \label{s-discuss-cf-wavebands}
In Fig.~\ref{f-KQVW}, the median standard deviation $s_i$ is most
noisy in the W band and a little noisy in the V band.
The quantitative best fit minima shown in Table~\ref{t-KQVW}
significantly favour the $-25.6$~ms offset hypothesis and 
reject the other hypotheses. The choice of intervals for 
mapmaking was optimised 
\prerefereechanges{for testing the $\delta t =0$ and 
$\delta t=0.5$ hypotheses rather than an offset in ms, which is why} the spreads of $\delta t \delta T$ for
V and W are less than for K, Ka and Q. Thus, with hindsight, given
the results found here, a {possible}
explanation for the higher noisiness in V and W 
{would be} that calculations
were not spread over a wide enough interval in $\delta t$
{in these bands}.

Another consistency check between
results in different wavebands is that
if the increase in $s_i$ away from a preferred timing offset
is mainly an effect of incorrect dipole-based calibration of
the observed dipole signal, then this dependence should only be
weakly dependent on waveband.

Figure~\ref{f-varmuKms} shows the same information as in
Fig.~\ref{f-KQVW}, with zeropoint removal but no scaling of the
standard deviations, i.e. $\sigma_i - A_i$, against timing offsets in 
milliseconds centred on
the hypothesis preferred by the data, i.e.  $(\delta t-0.5)\Delta t =
-25.6$~ms. In the range of approximately $(\delta t-0.5)\Delta t \in
(-25.6 \pm 175)$~ms, the V and W band results appear fully consistent
with the K and Ka, and Q results. The V and W bands do not appear
{to be} more noisy than the K and Ka, and Q bands. Moreover, the
dependence of the effect on $(\delta t-0.5)\Delta t$ appears to be
approximately independent of waveband, at least in this central range.
{\em A posteriori} calculation of maps
in W for larger absolute values of $\delta t$ (e.g. to about $\pm 640$~ms
as in the K and Ka bands) could reasonably
be expected to lead to a smaller uncertainty in this band.

\teighty

{This is indeed the case. Extra maps were calculated
for $\delta t = -10, \ldots, -6, 6, \ldots, 10$ in the twelve W band DA/year
combinations. If all the W maps are analysed together, the resulting best estimates 
are $\mu_{1/2}( \tau_i ) = 0.19 \pm 2.24$ and
$\mu_{1/2}( \tau_i-0.5 ) \Delta t = -16.1 \pm 116.8$~ms. 
However, Fig.~\ref{f-Wpmten} shows that the dependence of $s_i$ on $\delta t$ in the W band
is more complex than the quadratic dependence that gives visually acceptable fits in the 
K and Ka, Q, and V bands.}

{Figure~\ref{f-varmuKms} also shows that}
beyond the central range where the different bands appear consistent, 
there {appear to be significant differences
in the amplitude of the variation in $s_i$
between the different} 
wavebands, given the uncertainties estimated among DA/year
combinations within each waveband,
{even though the estimates of minima are clearly
consistent (Table~\ref{t-KQVW}).}
{This variation among wavebands might} be due to an increased
role of non-dipole components of the signal, e.g. Galactic foregrounds
that are brightest in the lower frequencies.  Since cosmological
perturbations should be much weaker than both the dipole and
low frequency foregrounds, the approximate frequency independence of the effect near
the minimum variance and frequency dependence away from the minimum is
qualitatively consistent with the effect near the minimum being
dominated by the dipole, and the effect further from the minimum
involving coupling effects between the dipole calibration,
the measured dipole, and foregrounds. 
{The beam size and sidelobe effects reported
by \nocite{SawShanks10a}{Sawangwit} \& {Shanks} (\protect\hyperlink{hypertarget:SawShanks10a}{2010}) and \nocite{LL10sidelobes}{Liu} \& {Li} (\protect\hyperlink{hypertarget:LL10sidelobes}{2010b}) may be related to this
latter speculation.}


\subsection{Sidelobe and other beam effects} \label{s-disc-sidelobe}
{\nocite{SawShanks10a}{Sawangwit} \& {Shanks} (\protect\hyperlink{hypertarget:SawShanks10a}{2010}) have estimated that WMAP beam profiles
at faint flux density levels, 
especially in the W band and weakly in the V band,
are wider than those estimated by the WMAP collaboration for bright flux density levels.
\nocite{LL10sidelobes}{Liu} \& {Li} (\protect\hyperlink{hypertarget:LL10sidelobes}{2010b}) have suggested that WMAP sidelobe interactions with
the dipole signal create an
effect that may be effectively modelled as if it were a positional
offset. 
From Fig.~1 of \nocite{WMAP1beam}{Page} {et~al.} (\protect\hyperlink{hypertarget:WMAP1beam}{2003}) and Fig.~2 of \nocite{WMAP5beam}{Hill} {et~al.} (\protect\hyperlink{hypertarget:WMAP5beam}{2009}), 
for the first year and 5-year WMAP data releases respectively,
it is clear that 
sidelobes in the beam shapes are strong for the W
DA's, weak for the V DA's, and much weaker for the K, Ka, and Q DA's.}

{The beam shape dependence on waveband
suggests an explanation not only for the variation in 
$s_i$
dependence on $(\delta t -0.5)\Delta t$ among wavebands at large 
 $|(\delta t -0.5)\Delta t|$, but also for the 
complexity in the shape of $s_i(\delta t)$ 
in the W band
(Fig.~\ref{f-Wpmten}) 
and the associated
large uncertainty in $\mu_{1/2}(\tau_i-0.5)\Delta t$ in the W band 
(Table~\ref{t-KQVW}).
The (partial) invertibility of a timing-offset--induced calibration as
presented in this work depends on the details of the scanning pattern. 
For a complicated beam profile, the relation between the sidelobe
orientation, the scanning pattern, and a slightly displaced dipole,
is likely to be complicated.  For a simple profile, the relation is
likely to be simpler, i.e. less noisy.
Thus, the 
complexity in $s_i(\delta t)$ 
in the W band is qualitatively consistent with 
\nocite{LL10sidelobes}{Liu} \& {Li} (\protect\hyperlink{hypertarget:LL10sidelobes}{2010b})'s estimate that sidelobe interaction with
the dipole is important in mapmaking, given 
the WMAP collaboration's estimates 
of the beam profiles shown in 
Fig.~1 of \nocite{WMAP1beam}{Page} {et~al.} (\protect\hyperlink{hypertarget:WMAP1beam}{2003}) and Fig.~2 of \nocite{WMAP5beam}{Hill} {et~al.} (\protect\hyperlink{hypertarget:WMAP5beam}{2009}). 
It is also
qualitatively consistent with the result of \nocite{SawShanks10a}{Sawangwit} \& {Shanks} (\protect\hyperlink{hypertarget:SawShanks10a}{2010}) that the 
difference between their radio-source--based beam profile estimates and 
the WMAP collaboration's Jupiter-based estimates is strongest in the W band.}

\subsection{Extra term from iterated maps} \label{s-disc-manyiters}
\prerefereechanges{
  Although use of an iterated map is equivalent to adding an extra
  term to Eqs~(\ref{e-1moment-defn}) and (\ref{e-2moment-defn}),
  introducing more noise into the method, the result should be
  statistically compatible with the above results.
  Checking the bands with fewer numbers of DA's and observations,
  i.e. adding the 80-th map iteration to these equations, 
  took about a calendar month of calculations.
  The results are shown in Table~\ref{t-eighty}.
  These mainly consist of an increase in the noise, with a slight
  increase in $\delta t$ values, i.e. 
  $\mu_{1/2} (\tau_i - 0.5) \Delta t = -20 \pm 6$~ms.
  This estimate still disagrees with $\delta t =0$ to high significance,
  and agrees with $(\delta t -0.5) \Delta t = -25.6$~ms at $0.8\sigma$.
}

\subsection{Devil's advocate: could the calibration be correct?}
\label{s-disc-alternatives}

It is difficult to see how the variance per pixel can be {\em decreased} by 
an arbitrary shuffling of the TOD.
The simplest explanation for the best estimate of
\prerefereechanges{$\mu_{1/2}(\tau_i -0.5) \Delta t = -29.7 \pm 2.5$~ms} is that the $-25.6$~ms
timing offset correction was not implemented when the WMAP
uncalibrated TOD were calibrated. 
However, could it be possible to avoid this
inference, i.e. could the calibration be correct?  Some possibilities
and counterarguments are as follows.

\begin{list}{(\roman{enumi})}{\usecounter{enumi}}
\item 
  As mentioned above, a model in which the dependence of $s_i$ on
  $\delta t$ is asymmetric might, in principle, give the minimum
  variance per pixel to occur near $\tau_i = 0.5$.  This would require
  a heuristic model for the asymmetric behaviour, and at least
  qualitatively would seem to require a much more complex shape than a 
  parabola.
  \begin{list}{(\roman{enumi}.\arabic{enumii})}{\usecounter{enumii}}
  \item
    In this case, it would be a coincidence that the symmetry
    assumption gives a result consistent with one of the two versions
    of \nocite{LL10toffset}{Liu} {et~al.} (\protect\hyperlink{hypertarget:LL10toffset}{2010a})'s hypothesis, as listed in
    Table~\ref{t-hypothesis}. That is, the successful prediction
    of one of the two versions of \nocite{LL10toffset}{Liu} {et~al.} (\protect\hyperlink{hypertarget:LL10toffset}{2010a})'s hypothesis 
    would be a coincidence.
  \end{list}
\item
  A heuristic model could relate to the fact that the K, Ka,
  and Q bands have much stronger Galactic contamination than the V
  (and W) bands. Could 
  \prerefereechanges{the $\mu_{1/2}(\tau_i -0.5) \Delta t = -29.7
    \pm 2.5$~ms} estimate be primarily an effect of Galactic contamination?
  \begin{list}{(\roman{enumi}.\arabic{enumii})}{\usecounter{enumii}}
  \item
    \prerefereechanges{The dependence of Galactic contamination on waveband is very
      strong going from K to W.  Tables~\ref{t-KQVW} and
      Table~\ref{t-domainshift} do not show any obvious trend in this
      sense.}
  \item 
    The statistic $s_i$ is a normalisation of $\sigma_i$, which is the
    {\em median}, over the pixels of a given map, of the standard
    deviation per pixel.  For Galactic contamination to affect the
    median, by shifting flux density onto or off given pixels, it
    would need to affect about half of the pixels at high Galactic
    latitude.  However, \prerefereechanges{a $-25.6$~ms} timing offset corresponds to only
    4{\arcm}, while pixel sizes are about 7.3$\ddeg$, i.e. about four
    orders of magnitude greater in solid angle. Even in the K and Ka
    bands, where the Galactic contamination is very strong, the number
    of tiny positional shifts of flux density onto or off a given
    pixel should be very high, so that the median over all valid
    pixels could be expected to vary smoothly and symmetrically around the
    correct timing offset.
  \item
    The coincidence argument also applies here. The Galactic
    contamination would have to mimic one of the two versions of
    \nocite{LL10toffset}{Liu} {et~al.} (\protect\hyperlink{hypertarget:LL10toffset}{2010a})'s hypothesis by chance, in sign 
    ($\tau_i < 0$), in amplitude, and in having an approximately
    constant value expressed as $\mu_{1/2}(\tau_i -0.5) \Delta t$.
  \end{list}
\item
  Could the small-scale blurring effect \nocite{Rouk10Toffa}({Roukema} \protect\hyperlink{hypertarget:Rouk10Toffa}{2010}, and references
    therein) have played a significant role in this
  analysis?
  \begin{list}{(\roman{enumi}.\arabic{enumii})}{\usecounter{enumii}}
  \item
    If it did, then it would have caused the results to be biased in
    favour of $\tau_i=0.5$, i.e.  $\mu_{1/2}( \tau_i-0.5 ) \Delta t =
    0$~ms. In other words, correcting the present result for the small-scale
    blurring effect would 
    \prerefereechanges{shift it in the opposite direction, giving 
      $\mu_{1/2}(\tau_i -0.5) \Delta t < -29.7
      \pm 2.5$~ms.} This would worsen the rejection of the $\tau_i=0.5$
    hypothesis.
  \end{list}
\item \prerefereechanges{\nocite{LL09lowquad}{Liu} \& {Li} (\protect\hyperlink{hypertarget:LL09lowquad}{2010a}) showed and
  \nocite{MSS10}{Moss} {et~al.} (\protect\hyperlink{hypertarget:MSS10}{2010}) confirmed that the CMB quadrupole is reduced in
  amplitude for $\Delta t(\delta t -0.5) \sim -25.6$~ms. If this is
  just a coincidence that relates the CMB quadrupole, the velocity
  dipole and the WMAP spacecraft scan pattern, could the variance
  minimisation method be strongly biased towards minimisation of the
  quadrupole, so that these constitute two instances of a single
  coincidence?  For example, let us suppose that $\delta t =0.5$ is
  the correct value. It is already known that in this case, the mean
  signal at $\delta t =0$ will be smaller.  Could this imply that
  the standard deviation per pixel at $\delta t =0 $ would be
  smaller too?
  \begin{list}{(\roman{enumi}.\arabic{enumii})}{\usecounter{enumii}}
  \item
    If $\delta t =0.5$ were really correct, then, although the {\em
      mean} signal at $\delta t = 0$ will be smaller, this would only be a
    result of the fortuitous cancelling, {\em in the mean,} of a component of the
    real signal by the dipole-induced difference signal. 
    However, reducing the mean by the addition of
    anticorrelated errors does not reduce the standard deviation per pixel.
    The scatter induced by a wrong assumed value of $\delta t$ is not
    cancelled by a reduced mean value.
  \item
    Calibration errors induced by a wrong value of $\delta t$ 
    apply to the full CMB signal, not just to the quadrupole part of the
    signal. The latter is a very weak part of the full CMB signal. Thus,
    scatter induced by a wrong value of $\delta t$ can have
    only a weak dependence on the quadrupole. Table~\ref{t-KQVW} shows
    that $\delta t = 0.5$ is still strongly rejected in the K and Ka, 
    and Q bands, where galactic foregrounds play a strong role.
  \end{list}
}
\end{list}

Retaining the hypothesis that the calibration was correct does not
seem to be easy.

\section{Conclusion} \label{s-conclu}

While the $-25.6$~ms offset between the times in the Meta Data Set
and the Science Data Set in the WMAP TOD files, discovered by
\nocite{LL10toffset}{Liu} {et~al.} (\protect\hyperlink{hypertarget:LL10toffset}{2010a}), did not lead to an error in compiling the
calibrated, filtered, 3-year TOD into maps \nocite{Rouk10Toffa}({Roukema} \protect\hyperlink{hypertarget:Rouk10Toffa}{2010}), it is
difficult to avoid the conclusion that this timing offset {\em did}
induce an error at the calibration step. 
The two different versions of 
\nocite{LL10toffset}{Liu} {et~al.} (\protect\hyperlink{hypertarget:LL10toffset}{2010a})'s hypothesis
gave numerical predictions, listed in Table~\ref{t-hypothesis}.
The constant offset \prerefereechanges{in ms} hypothesis is consistent with \prerefereechanges{the} analysis
of the variance maps, summarised in Table~\ref{t-KQVW}.
\prerefereechanges{This result does not seem to be sensitive to the fitting method
(\SSS\ref{s-disc-fit-sensitivity}),
it is consistent among all wavebands 
(\SSS\ref{s-discuss-cf-wavebands}), 
using highly iterated maps adds noise and
gives consistent results (\SSS\ref{s-disc-manyiters}),
and 
alternative explanations seem to be speculative and/or fail to save the
$\delta t = 0.5$ hypothesis (\SSS\ref{s-disc-alternatives}).}
The timing offset that is
numerically implicit in the TOD files (e.g., see Appendix A,
\nocite{Rouk10Toffa}{Roukema} \protect\hyperlink{hypertarget:Rouk10Toffa}{2010}) would appear to provide the simplest
explanation.
\prerefereechanges{A new analysis of 7 years of Q, V, and W band WMAP TOD by \nocite{LXLWMAP7}{Liu} {et~al.} (\protect\hyperlink{hypertarget:LXLWMAP7}{2010b}) 
finds a similar result.}
\prerefereechanges{Thus, it appears 
  that (1) in the calibration step, both the
  spacecraft attitude quaternion timestamps and the observational
  timestamps were used, for the dipole and observational data,
  respectively, inducing a calibration error, while (2) in the mapmaking
  step, the observational timestamps were (correctly) assumed to be
  correct both for dipole and observational data, inducing no further
  error, but retaining the original calibration error.}
This supports the claims by \nocite{LL10toffset}{Liu} {et~al.} \nocite{LL09lowquad,LL10toffset}({Liu} \& {Li} \protect\hyperlink{hypertarget:LL09lowquad}{2010a}; {Liu} {et~al.} \protect\hyperlink{hypertarget:LL10toffset}{2010a})
that the CMB quadrupole has been substantially overestimated.
A rough estimate \prerefereechanges{made here} 
is that the quadrupole estimated in maps based on the
wrongly calibrated TOD is overestimated by about 
\prerefereechanges{$64\pm6\%$ to $94\pm10\%$}
for the KQ85 and KQ75 sky masks respectively (\SSS\ref{s-disc-quad}).

\begin{acknowledgements}
Thank you to Hao Liu and Ti-Pei Li for useful public and private
discussion and making their software publicly available, and to 
Bartosz Lew, Roman Feiler, Martin France, \prerefereechanges{Tom Shanks
and an anonymous referee}
for useful comments.
Use was
made of the WMAP data (\url{http://lambda.gsfc.nasa.gov/product/}),
the {\sc GNU Data Language}
(\url{http://gnudatalanguage.sourceforge.net}), 
the WMAP IDL{\textregistered}/GDL routines 
(\href{http://lambda.gsfc.nasa.gov/data/map/dr4/software/widl_v4/wmap_IDLpro_v40.tar.gz}{{\tt http://lambda.gsfc.nasa.gov/}}
\href{http://lambda.gsfc.nasa.gov/data/map/dr4/software/widl_v4/wmap_IDLpro_v40.tar.gz}{{\tt data/map/dr4/}}\href{http://lambda.gsfc.nasa.gov/data/map/dr4/software/widl_v4/wmap_IDLpro_v40.tar.gz}{{\tt software/widl\_v4/wmap\_IDLpro\_v40.tar.gz}}),
the {IDL{\textregistered} Astronomy User's Library}
(\url{http://idlastro.gsfc.nasa.gov}),
the utility program {\sc HPXcvt}
from the {\sc WCSLIB} library
(\url{http://www.atnf.csiro.au/people/mcalabre/WCS/}),
the Centre de Donn\'ees astronomiques de Strasbourg
(\url{http://cdsads.u-strasbg.fr}),
and
the GNU {\sc Octave} command-line, high-level numerical computation software 
(\url{http://www.gnu.org/software/octave}).

%
%
%

\end{acknowledgements}

\subm{ \clearpage }

\nice{
%

}


\end{document}